\documentclass [prb,twocolumn,notitlepage,superscriptaddress,floatfix] {revtex4-1}
\usepackage{amsmath}
\usepackage{graphicx}
\usepackage{lmodern}
\usepackage{amsmath}

\usepackage{color}
\usepackage{amssymb}
\newcommand{\beq} {\begin{equation}}
\newcommand{\eeq} {\end{equation}}
\newcommand{\bea} {\begin{eqnarray}}
\newcommand{\eea} {\end{eqnarray}}
\newcommand{\be} {\begin{equation}}
\newcommand{\ee} {\end{equation}}
\renewcommand{\(}{\left(}
\renewcommand{\)}{\right)}
\renewcommand{\[}{\left[}
\renewcommand{\]}{\right]}

\DeclareMathOperator{\sgn}{sgn}

\begin{document}

\title {Superconductivity near a quantum-critical point: The special role of the first Matsubara frequency}
\author{Yuxuan Wang}
\affiliation{Department of Physics and Institute for Condensed Matter Theory, University of Illinois at Urbana-Champaign, Urbana, Illinois 61801, USA }
\author{Artem Abanov}
\affiliation{Department of Physics. Texas A\&M University, College Station,  USA}
\author{Boris L. Altshuler}
\affiliation{Department of Physics. Columbia University, New York,  USA}
\author{Emil A. Yuzbashyan}
\affiliation{Center for Materials Theory, Department of Physics and Astronomy, Rutgers University, Piscataway, NJ 08854, USA}
\author{Andrey V. Chubukov}
\affiliation{School of Physics and Astronomy and William I. Fine Theoretical Physics Institute,
University of Minnesota, Minneapolis, MN 55455, USA}
\date{\today}

\begin{abstract}
 Near a quantum-critical point in a metal
 strong fermion-fermion interaction mediated by a soft collective boson
 gives rise to incoherent, non-Fermi liquid behavior.
It
also often gives rise to
 superconductivity which masks the non-Fermi liquid behavior.
  We analyze the interplay between
   the tendency to pairing
    and fermionic incoherence for a set of
 quantum-critical models with
 effective dynamical interaction between low-energy fermions.  We argue that
 superconducting $T_c$ is non-zero even
 for strong incoherence and/or weak interaction due to the fact that
the self-energy from dynamic critical fluctuations vanishes for the two lowest fermionic Matsubara frequencies $\omega_m = \pm \pi T$.
   We obtain the analytic formula for $T_c$ which  reproduces well earlier numerical results for the  electron-phonon model at
  vanishing Debye frequency.
\end{abstract}
\maketitle

{\it \bf Introduction.}~~~The interplay between superconductivity and non-Fermi liquid behavior in metals is one of most fascinating issues in
 the modern physics of correlated electron
 systems~\cite{combescot,nick_b,acf,acs,son,sslee,subir,moon_2,max,senthil,raghu,mack,scal,max_last,efetov,raghu_15,steve_sam,varma}
 A generic metallic system in $D >1$ is a Fermi liquid with coherent quasiparticles at low energies.  This coherence is destroyed if the system is brought  to a quantum-critical point (QCP),  beyond which it develops
  an  electronic order
   in spin or charge channel.  At a QCP  fluctuations of the order parameter become massless.
  In $D \leq 3$,
  the four-fermion interaction, mediated by these massless fluctuations, destroys fermionic coherence
   at $T=0$,
   either at specific hot points on the Fermi surface~\cite{acs,efetov,max_2,struck}, if the order has a finite momentum, or everywhere on the Fermi surface, if the order develops with $q=0$ (Ref.\ \onlinecite{q=0}).  The same  massless fluctuations, however,  also mediate the pairing interaction, and if this interaction has an attractive  angular component
    the system can develop a superconducting instability at a finite $T$, before a QCP is reached. A dome of superconductivity above a QCP
  prevents a non-Fermi liquid, QC behavior from
   extending
   down to the lowest energies.

 The existence of superconductivity near a QCP is not guaranteed, however, because strong fermionic self-energy acts  against pairing.
   There are two effects from the self-energy.  First,
    at
     $ T \neq 0$ the self-energy  from static (thermal)
     fluctuations acts as an impurity and may cause pair-breaking.
   This is crucial for spin-triplet superconductivity,
   %
   for which
    thermal self-energy acts as a magnetic impurity \cite{triplet},
   but not for  spin-singlet superconductivity,
    for which   it
    acts as a non-magnetic impurity and its singular contribution  cancels out
      by
       Anderson theorem~\cite{msv}.
   In  this paper we
     consider spin-singlet pairing and  neglect the contribution from thermal fluctuations.
   Second,
   already at $T =0$
   the self-energy  produces  strong upturn mass renormalization
     and
    shrinks  the range of a coherent  fermionic behavior.
     Both these effects are detrimental to superconductivity.
      %

   The pairing amplitude and the self-energy come from the same underlying interaction mediated by a soft boson, hence the two are generally
   of the same order.
    Zero-temperature studies
     of specific models in $D=2$ and in $D =3-\epsilon$
   have shown~\cite{acf,acs,son,subir,moon_2,senthil,max_last}  that superconductivity does develop at a QCP, however
      these
      studies also hinted~\cite{nick_b,acf,acs} that the pairing at a QCP is a threshold problem and may disappear if the self-energy
       gets enhanced compared to the pairing amplitude.   A recent study~\cite{raghu_15} made this explicit by  extending a model in $D=3-\epsilon$  to
         large $N$ in such a way  that the self-energy  gets enhanced, while the pairing amplitude remains intact.  The authors of Ref.\  \onlinecite{raghu_15}
          performed $T=0$ analysis  and
          argued that there exists a critical $N$  above which the pairing does not develop
            because decoherence, caused by strong self-energy, wins over the tendency to pairing due to  an attraction.

  In this communication we
  %
  and analyze the same pairing problem, but  at a non-zero $T$.
  Our result is different from  Ref.\ \onlinecite{raghu_15} and earlier work by some of us (Ref.\ \onlinecite{acf}) --
      we argue that superconducting
      $T_c$ is finite at arbitrary $N$.
     %
     The reason
    is that the competition between
     the self-energy and the pairing interaction at a finite temperature is qualitatively different from that at $T=0$.
      Namely, at a finite $T$ the Matsubara self-energy $\Sigma (\omega_n)$ is  a discrete variable, defined at a set of  $\omega_n = \pi T (2n+1)$. It still large for all $n \neq 0, -1$,
       but  at the  two lowest Matsubara frequencies $\omega_n = \pm \pi T$ it vanishes if we neglect the contribution from static bosonic fluctuations.~\cite{maslov}
     %
       At the same time, the pairing interaction $\chi (\Omega_m)$,  also taken without the static part (i.e., at bosonic $ \Omega_m = 2 \pi T m, m \neq 0$)
        is not reduced at $\Omega_m = \pi T - (-\pi T) = 2\pi T$ compared to $\chi (\Omega_m)$ at  other $\Omega_m$.
           As a result,
       the pairing interaction between fermions with $\omega_n = \pm \pi T$  is strong, while the  competing  contribution from the self-energy is absent.
        Although this holds only for the two Matsubara frequencies, we show that this is sufficient to render $T_c$ finite. 
         Moreover,
           $T_c$ is not small and has a power-law dependence of the coupling constant, which is stronger than 
             than the logarithmical divergence in BCS theory,  although the latter is obtained by summing up an infinite set of Matsubara points.

In broader terms, we argue against the commonly used procedure\cite{acf,acs,subir,moon_2,senthil,max_last,raghu_15} to obtain $T_c$ at a QCP by
   computing the pairing susceptibility  $\chi_{pp} (\omega)$ at $T=0$, associating the superconducting region with the range of $N$ where $\chi_{pp} (\omega)$
   becomes negative below some $\omega^*$,  and identifying $T_c$ with $O(\omega^*)$.  We argue that $T_c$ has to be determined from the actual calculations at a finite $T$, and $T_c$  generally does not scale with $\omega^*$, except for special cases like models in $D =3-\epsilon$ and $N= O(1)$.

To be specific, our conclusion holds for a set of QC models with dynamical interaction between fermions, for which  the Eliashberg approximation~\cite{eliashberg} is valid.
Within
 this
  approximation, the momentum integration in the gap equation can be carried out explicitly,
 and the analysis of
  superconductivity reduces to a set of equations for the frequency dependent pairing vertex $\Phi (\omega_m)$ and fermionic self-energy $\Sigma (\omega_m)$,
   both originating from the effective,
   momentum-averaged interaction $\chi (\omega_m - \omega'_m)$.   We consider a generic case of  $\chi (\Omega_m) = (g/|\Omega_m|)^\gamma$, where $g$ is the effective fermion-boson coupling.  We list specific examples of
     different
     $\gamma$  below. In particular, $\gamma =2$ corresponds to much studied strong coupling limit of electron-phonon interaction~\cite{combescot,others,phon_rev,ital}. We argue that  $T_c$ is non-zero for any $\gamma$, even if the self-energy is enhanced
  after a proper extension of the model to large $N$,
   as in \cite{raghu_15}.
    Moreover, at large $N$,
    $T_c \approx [g/(2\pi)]/ N^{1/\gamma} \approx 0.16 g/ N^{1/\gamma}$
is fully determined by the two lowest Matsubara frequencies.
 At
 $N=1$
 this formula
  yields
  $T_c \approx 0.16 g$.
   This value
  is very close to
 $T_c \approx 0.18 g$ obtained numerically for $\gamma =2$ (Refs.\ \onlinecite{ad,others}).
   which
    implies that $T_c$ for QC electron-phonon problem is predominantly determined by just the two lowest Matsubara frequencies.

{\it \bf The model.}~~~ We consider a system of fermions at the boundary between a Fermi liquid state and a state with a long-range order in either spin or charge channel (ferromagnetism, nematic order,
spin/charge-density-wave, etc).  At a QCP, the propagator of a soft boson
 becomes massless and
 mediates singular interaction between fermions.
 Like we said, we treat this interaction as attractive in at least one pairing channel. This is true for QCP towards density-wave instabilities~\cite{kl}, but we caution that
  this is not always the case -- e.g., for fermions at the half-filled lowest Landau level, long range current-current interaction mediated by  gapless gauge fluctuations is repulsive in all channels~\cite{max_last}.

 We assume, following earlier work~\cite{acf,acs,max,senthil,scal,efetov,max_last,raghu_15,migdal,haslinger}  that bosons can be treated as slow modes compared to fermions, i.e.,  the Eliashberg approximation is valid.
Within this approiximation   one can explicitly integrate over the momentum component perpendicular to the Fermi surface reduce the integral equations for the self-energy $\Sigma$
   and the pairing vertex $\Phi$ to the set for $\Sigma ({\bf k}_F, \omega_m)$ and $\Phi ({\bf k}_F, \omega_m)$ on the Fermi surface.
   We will be interested in the solution for $T_c$, hence we set $\Phi ({\bf k}_F, \omega_m)$ to be infinitesimally small and approximate  $\Sigma ({\bf k}_F, \omega_m)$ by its normal state value.
     We make one additional approximation -- assume that the dependence of $\Phi ({\bf k}_F, \omega_m)$ on $\omega_m$ and on
     on  the momentum
     direction along the Fermi surface can be factorized, i.e. that
    $\Phi ({\bf k}_F, \omega_m) = f_\Phi ({\bf k}_F) \Phi (\omega_m)$, where
     $f_\Phi$ has the symmetry of the corresponding superconducting state~\cite{acs,varma}, and neglect the momentum dependence of $\Sigma ({\bf k}_F, \omega_m)$.
       Under this approximation, the integration over momentum component along the Fermi surface can be done explicitly~\cite{acs,efetov},
     and the set of equations for
      $T_c$ reduces to the integral equation for $\Phi (\omega_m)$ and the equation for the normal state self-energy $\Sigma (\omega_m)$:
    \bea
    &&\Phi (\omega_m) =
    \frac{g^\gamma}{N}  \pi T \sum_{m' \neq m} \frac{\Phi (\omega_{m'})}{|\omega_{m'} + \Sigma (\omega_{m'})|} ~\frac{1}{|\omega_m - \omega_{m'}|^\gamma}, \nonumber \\
    && \Sigma (\omega_m) = g^\gamma \pi T \sum_{m' \neq m}  \frac{{\text{sign}}(\omega_{m'})}{|\omega_m - \omega_{m'}|^\gamma},
\label{1}
\eea
where we incorporated the overall factors from the integration over momentum into $g$.
Like we said, we
 neglect
 the terms with $m = m'$ in Eq.\ (\ref{1})
  because for spin-singlet pairing such terms
   cancel out between $\Phi (\omega_m)$ and $\Sigma (\omega_m)$
   We discuss this in more detail in Ref.\ \onlinecite{SM}.
    The overall factor  $1/N$ is the result of  extending the model to an SU($N$) global symmetry which involves both fermions and bosons\cite{raghu_15}. We
 treat $N$ as a parameter.  Our goal is to understand  whether there is a critical  $N$ above which $T_c =0$, i.e. the normal state extends down to $T=0$.

Models described by Eq. (\ref{1}) include a model for color superconductivity~\cite{son} ($\gamma = 0_+$, $\chi (\Omega_m) \propto \log{|\omega_m|}$),  models for spin- and charge-mediated pairing in $D=3-\epsilon$ dimension~\cite{senthil,max_last,raghu_15} ($\gamma = O(\epsilon) \ll 1$),  a 2D pairing model~\cite{2kf}  with  interaction peaked at $2k_F$ ($\gamma =1/4$),  2D models for  pairing at a nematic/Ising-ferromagnetic QCP~\cite{nick_b,steve_sam,triplet} ($\gamma =1/3$),  a 2D hot-spot model for pairing at the $(\pi,\pi)$ SDW QCP~\cite{acf,acs,millis_05,wang}
 and at a 2D CDW QCP~\cite{ital,wang_2}, 2D models for  pairing  by undamped fermions ($\gamma =1$), the strong coupling limit of phonon-mediated superconductivity~\cite{combescot,ad,others}, and  models with parameter-dependent $\gamma$ (Refs. \onlinecite{subir,moon_2}).

   {\it \bf  The argument for the threshold.}~~~ To set the stage for our analysis, we briefly display the argument for the existence of a threshold in $N$ for $T_c$.
The argument is based on  the  analysis of the pairing susceptibility  at $T=0$ for $0 < \gamma <1$ (Refs.\  \onlinecite{acf,raghu_15}).
 At $T=0$ the self-energy has a non-Fermi liquid form:  $\Sigma (\omega_m) = |\omega_m|^{1-\gamma} \omega^\gamma_0  {\text{sign}} (\omega_m)$, where $\omega_0 = g [2/(1-\gamma)]^{1/\gamma}$.  Substituting this $\Sigma (\omega_m)$ into the equation for $\Phi (\omega)$ and adding up a bare $\Phi_0$, one can compute the $T=0$ pairing susceptibility $\chi_{pp} (\omega_m) = \Phi (\Omega_m)/\Phi_0$ at $\omega_m < \omega_0$  order by order in $1/N$.
 The building block for series for $\chi_{pp} (\omega_m)$ is $\int d \omega_{m'} 1/(|\omega_{m'} - \omega_{m}|^\gamma |\omega_{m'}|^{1-\gamma})$, where
   the first term comes from the interaction and the second from the self-energy. The integrand scales as $ 1/|\omega_{m'}|$ at
   $\omega_{m'} > \omega_m$,  hence the series for $\chi_{pp} (\omega_m)$ are logarithmical.
     At $N \gg 1$
      the coupling is weak and
        one can just  sum up the series of leading logarithms,
         like in BCS theory. However, this analogy does not go further
         because in
          our case,  each logarithm is cut  by $\omega_m$ rather than by $T$  and  the summation of the logarithms yields $\chi_{pp} (\omega_m) = 1 + \alpha \log{\frac{\omega_0}{|\omega_m|}} + \frac{\alpha^2}{2} \left(\log{\frac{\omega_0}{|\omega_m|}}\right)^2 +
\frac{\alpha^3}{6} \left(\log {\frac{\omega_0}{|\omega_m|}}\right)^3 + ... =
\left(\omega_0/|\omega_m|\right)^\alpha$,
 where $\alpha =(1-\gamma)/N$.  This susceptibility is positive
 i.e.
  the summation of the leading logarithms does not give rise to  pairing.

\begin{figure}
\includegraphics[width=\columnwidth]{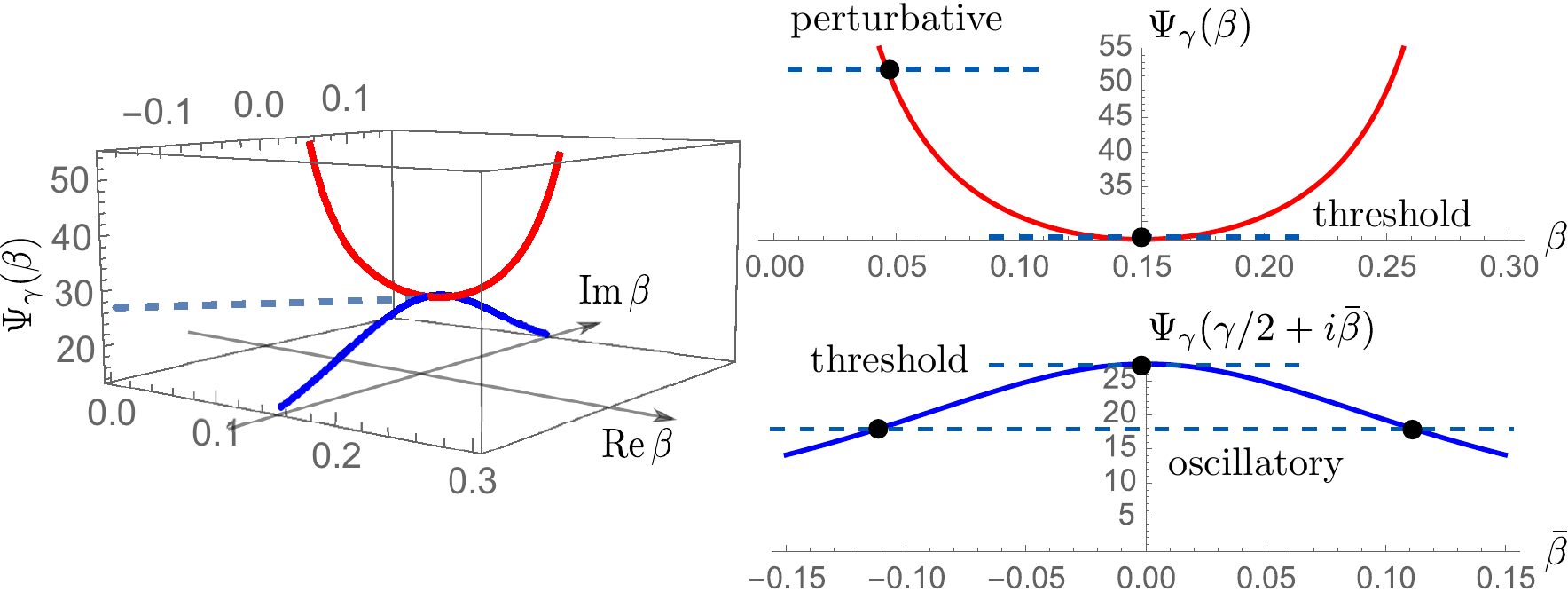}
\caption{Left: The plot of the function $\Psi_\gamma (\beta)$
 for $\gamma =0.3$.
 Right: the solution of the equation $((1-\gamma)/2N) \Psi_{\gamma} (\beta) =1$. For large $N$, $\beta$ is real, as in perturbation theory (red line), i.e., superconductivity does not develop.  For $N < N_{cr}$,
  $\beta = \gamma/2 \pm i {\bar \beta}$ is complex (blue line). For a complex $\beta$, $\Phi (\omega_m)$ is oscillatory in frequency, implying that $T_c$ is finite.
  }

\label{fig_0}
\end{figure}
  This line of reasoning is developed further by  solving for the susceptibility beyond the logarithmical approximation.
  The $1/\omega_{m'}$ scaling of the kernel suggests  a power-law form $\Phi (\omega_m) \propto (\omega_0/|\omega_m|)^{\beta}$  at
   $\omega_m < \omega_0$.
  Substituting this into (\ref{1}) and evaluating the integrals, we obtain an equation on ${\beta}$ of the form $(\alpha/2) \Psi_{\gamma} (\beta) =1$, where $  \Psi_{\gamma} (\beta) = \frac{\Gamma({\beta}) \Gamma(\gamma-{\beta})}{\Gamma(\gamma)} + \Gamma(1-\gamma)\left(\frac{\Gamma({\beta})}{\Gamma(1-\gamma + {\beta})} + \frac{\Gamma(\gamma-{\beta})}{\Gamma(1- {\beta})} \right) \nonumber.$
 We plot $\Psi_{\gamma} (\beta)$ in Fig.\ \ref{fig_0}.
Solving for ${\beta}$ as a function of $\alpha$ and $\gamma$ and
choosing the branch which gives ${\beta} \approx  \alpha$ at small $\alpha$,
 consistent with logarithmical perturbation theory,
 we find
    that ${\beta}$
   increases with  $\alpha$,
    reaches  the value $\gamma/2$ at a critical $\alpha_{cr} = (1-\gamma)/N_{cr}$,
     and at larger $\alpha$ (i.e., smaller $N$)  becomes complex:
  ${\beta} = \gamma/2 \pm i {\bar \beta}$, where ${\bar \beta} \propto
   (\alpha- \alpha_{cr})^{1/2} \sim (N_{cr} - N)^{1/2}$. As the consequence, $\chi_{pp}$ becomes an oscillating function of $\omega_m$: $\chi_{pp} (\omega_m)  \propto (\omega_0/|\omega_m|^{\gamma/2}) \cos({\bar \beta} \log(\omega_0/|\omega_m|) + \psi_0)$, where $\psi_0$ is an arbitrary
   phase.
   Oscillations of the pairing susceptibility cannot be obtained within a perturbation theory
     and their presence
      was interpreted as the sign
      that the system has already underwent  a pairing instability at some finite $T_c$.
       To obtain $T_c$, earlier works used $T=0$ form of $\chi_{pp} (\omega_m)$
     and identified $T_c$ with the largest $\omega_m$ at which $\chi_{pp} (\omega_m)$ first becomes negative.
          At $\alpha \geq \alpha_{cr}$, when $\beta$
     is small,  this yields~\cite{acf,raghu_15,khvesh}
     $T_c \sim \omega_0 e^{- a/(N_{cr} - N)^{1/2}}$, where $a = O(1)$.

   {\it \bf Finite $T$ analysis.}  We now perform   the actual analysis at a finite $T$ and argue
    that it yields a result different from the one at $T=0$. Namely, we argue that $T_c$ is non-zero for any $N$ and
    only tends to zero when $N$ tends to infinity.
     We show that this result originates from the vanishing of the self-energy at Matsubara frequencies $\omega_m = \pm \pi T$. The special role of the lowest
      Matsubara  frequencies cannot be  detected in the  $T=0$ analysis
       in which
        Matsubara frequency is
         a continuous variable.

  Vanishing of the self-energy $\Sigma (\omega_m = \pm \pi T)$ can be readily seen from Eq.\ (\ref{1}). We have $\Sigma (\pi T) =  [g/(2\pi T)]^\gamma \pi T \sum_{m' \neq 0} {\rm sign} (2m'+1)/|m'|^\gamma$, and the sums over positive and negative $m'$ cancel each other.
   The same holds for $\omega_m = -\pi T$. For any other $m \geq 1$, $\Sigma (\omega_m>0) \sim \omega_m (g/(2\pi T))^\gamma \gg \omega_m$, i.e at low $T$ the self-energy at
     $|\omega_m| \neq \pi T$ well exceeds the bare $\omega_m$ term in the fermionic propagator.
   Note in passing that the vanishing of $\Sigma (\omega_m = \pm \pi T)$ in our analysis does not actually imply that at this frequency a fermion is a free quasiparticle, because we eliminated from $\Sigma (\omega_m)$ the contribution from static critical fluctuations (the $m=m'$ term in in Eq.\ (\ref{1})). Such contribution is irrelevant for the pairing, but it is parametrically larger than $T$ near a QCP, hence the full self-energy has a non-Fermi liquid form even at $\omega_m = \pm \pi T$.

To make our point
about $T_c$,
 we consider large $N$ and small $T$. Neglecting $\omega_m$ compared to the self-energy for all $m$ except  $m =0$ and $m=-1$, using the symmetry conditions
$\Phi (\omega_m) \equiv \Phi_m = \Phi_{-m-1}$ and $\Sigma (\omega_m) \equiv \Sigma_m =-\Sigma_{-m-1}$, and introducing ${\bar \Phi}_m\equiv\Phi_m /(\pi T K_T) ,~{\bar \Sigma}_m\equiv\Sigma_m /( \pi T K_T) $, where $K_T = [g/(2\pi T)]^\gamma \gg 1$,
 we re-write the gap equation in (\ref{1}) as a set of coupled equations for ${\bar \Phi}_{m=0,-1}$ and ${\bar \Phi}_{m>0}$:
\bea
&&{\bar \Phi}_0 =  \frac{K_T}{N} {\bar \Phi}_{-1} +\frac{1}{N} \sum_{m>0} \frac{{\bar \Phi}_m}{{\bar \Sigma}_m} \left[\frac{1}{m^\gamma} + \frac{1}{(m+1)^\gamma}\right]   \nonumber \\
&&  {\bar \Phi}_{m>0} = \frac{K_T}{N}  \left[\frac{{\bar \Phi}_0}{m^\gamma} + \frac{{\bar \Phi}_{-1}}{(m+1)^\gamma}\right]
 \nonumber \\ &&
+\frac{1}{N} \sum_{m'>0, m' \neq m} \frac{{\bar \Phi}_m}{{\bar \Sigma}_m} \left[\frac{1}{|m-m'|^\gamma} + \frac{1}{(m+m'+1)^\gamma}\right] \label{4}
 \eea
We distinguish ${\bar \Phi}_0$  and ${\bar \Phi}_{-1}$ in (\ref{4})   only for illustrative purposes. In fact,   the two are equal,
${\bar \Phi}_0 = {\bar \Phi}_{-1}$.

At vanishing $1/N$ Eq. (\ref{4}) has a solution at $K_T=N$, i.e. at $T = T_c = (g/2\pi)/N^{1/\gamma}$.
 Indeed, the first equation in (\ref{4}) is satisfied, while the second one determines
${\bar \Phi}_{m}$ for all $m>0$ in terms of ${\bar \Phi}_0$:
 ${\bar \Phi}_{m>0} = {\bar \Phi}_0 \left[\frac{1}{m^\gamma} + \frac{1}{(m+1)^\gamma}\right]$.  Plugging this  ${\bar \Phi}_{m>0}$ into the first equation in (\ref{4}),
   we
    obtain $T_c$ with $1/N$ correction  (see Ref. \onlinecite{SM} for details)
\beq
T_c \approx \frac{g}{2\pi} \frac{1}{N^{1/\gamma}} \left(1 + \frac{\delta_\gamma}{N \gamma}\right),
\label{6}
\eeq
where $\delta_\gamma=\sum_{m>0}{[1/m^\gamma+1/(m+1)^\gamma]^2}/\bar\Sigma_m$  is
 a number of order one.
 We see that $T_c$ is non-zero for {\it any} $N$, i.e., no matter how strong is the self-energy at Matsubara frequencies $\omega_m$ with $m \neq 0, -1$.
We also see  that  superconducting $T_c$
 is predominantly determined by
   the two lowest Matsubara frequencies, for which the pairing interaction is strong, but the self-energy vanishes.  This new understanding is very different from the previous one that superconductivity at a QCP
    originated  from the pairing of incoherent fermions at $T=0$.


 \begin{figure}
 \includegraphics[width=.9\columnwidth]{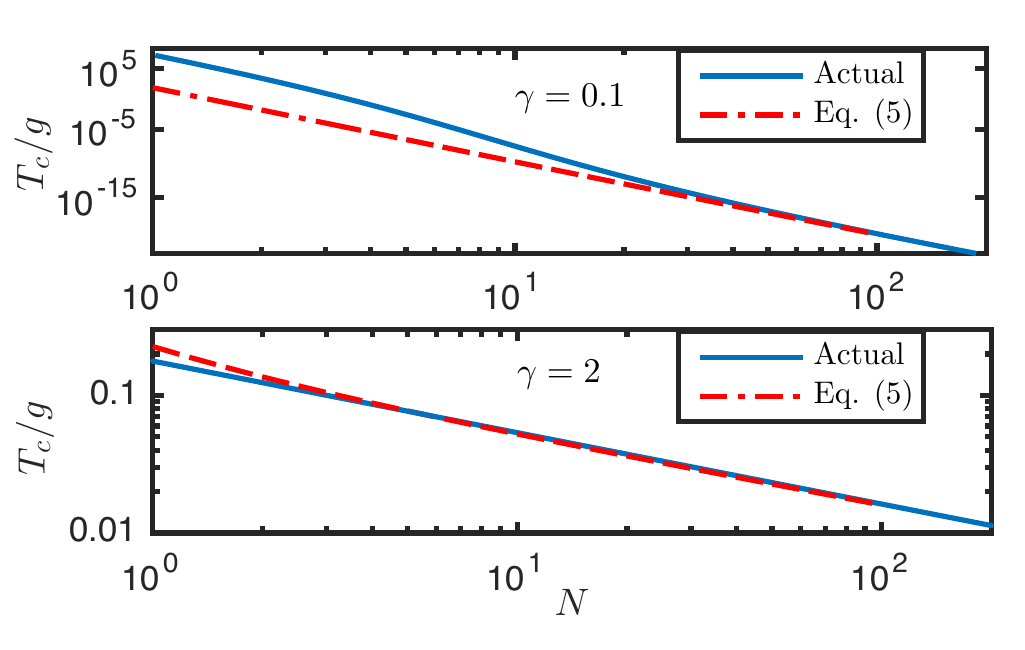}
 \caption{
  Superconducting $T_c$, obtained by   solving  the gap  Eq.~(\ref{4}) numerically (labeled as ``actual''),
   vs. the analytical result from Eq.\ (\ref{6}).  Upper panel: $\gamma=0.1$, lower panel: $\gamma =2$.
    In both cases  the analytical $T_c$ perfectly matches the numerical one at large $N$.  For $N = O(1)$, the numerical solution
    yields  much larger $T_c$ than Eq.~(\ref{6}) for $\gamma =0.1$, but for $\gamma =2$  numerical and analytical results  remain close even for $N=1$.}
 \label{fig_1}
 \end{figure}

 {\it \bf The value of $T_c$}. ~~~In Fig.\ \ref{fig_1} we show $T_c$ given by Eq.\ (\ref{6}), together with the numerical solution of the gap equation.
 We see that at large $N$ the actual solution and the one from Eq. (\ref{6}) agree quite well, as expected.  The agreement does not extend to $N \sim 1$ at small $\gamma$, but gets progressively better for larger $\gamma$, for which $T_c$ is predominately determined by the first two Matsubara frequencies even for $N =1$,
  i.e. $T_c \approx g/(2\pi)$.  Other Matsubara frequencies account only for a small correction to
  $T_c = g/(2\pi)$.  To verify this,  we computed the leading correction in $1/\gamma$ for an arbitrary $N$  and obtained
$  T_{c} = \frac{g}{2\pi} ({s}/{N})^{1/\gamma}$
   where $s = s(N)$ is determined from $J_{3/2+N/s} (1/s)/J_{1/2+N/s} (1/s) = s-1$, where $J$ is a Bessel function (see Ref. \cite{SM} for detail)
 At $N =1, s= 1.1843$, at $N \gg 1$, $s = 1 + 1/(2N)$,  in agreement with
    Eq.\ (\ref{6}) (In Eq.\ (\ref{6}), $\delta_\gamma \to 1/2$ at $\gamma\to \infty$).
    For the strong coupling limit of electron-phonon superconductivity ($\gamma =2$, $N =1$),
  $T_c \approx 0.17 g$, which is very close to $0.18 g$, obtained in extensive numerical studies~\cite{ad,others} on a large mesh of Matsubata frequencies.
   This
    has been noticed in Ref.\ \onlinecite{ad} but not related to the absence of the self-energy at $\omega_m = \pm \pi T$.

        \begin{figure}
    \includegraphics[width=0.9\columnwidth]{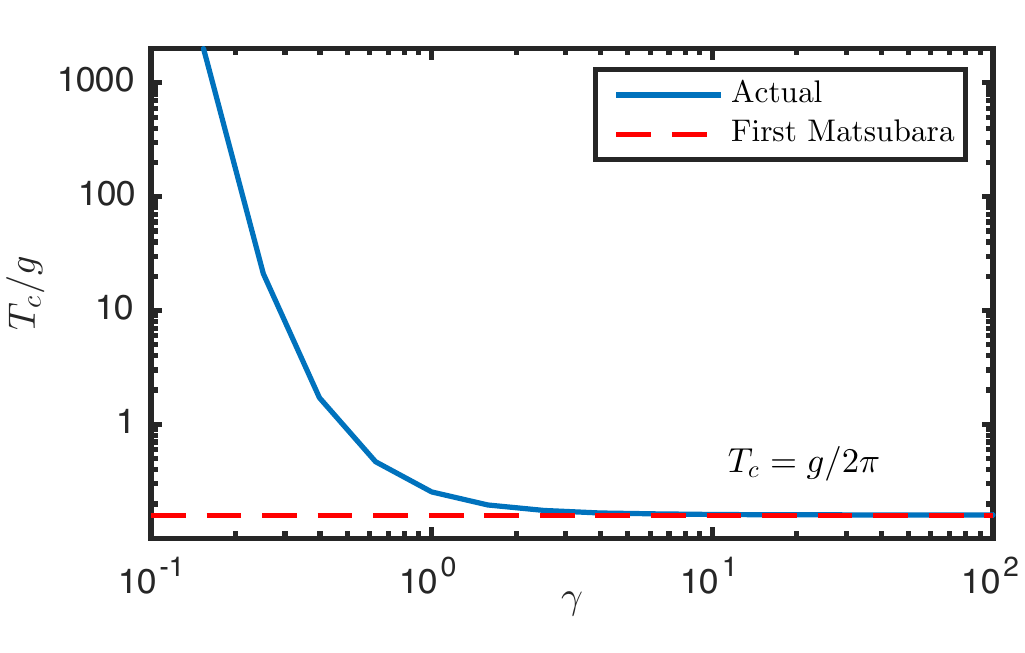}
    \caption{The numerical result for $T_c$ at $N=1$ as a function of $\gamma$. At small $\gamma$, $T_c$ is determined by all Matsubara frequencies and increases exponentially with decreasing $\gamma$ (see the text).  At $\gamma >1$ it rapidly approaches $T_c =g/2\pi$, which we obtained analytically from  the two lowest
     Matsubara frequencies.}
        \label{fig_2}
    \end{figure}

  For completeness, we also computed $T_c$ at small $\gamma$ and $N = O(1)$.
   In this regime
   %
 %
 $T_c \gg \omega_0$ (see Fig.\ \ref{fig_1}) and
  the self-energy is again irrelevant, but now simply because at $T= T_c$, $\omega_m \gg \Sigma (\omega_m)$  for all $m$.
 Neglecting $\Sigma (\omega_{m'})$ in
  Eq. (\ref{1}),
   we obtain (see Ref. \cite{SM} for details)
 \be
  T_c \sim \omega_0(\gamma N)^{-1/\gamma}  \sim  \frac{g}{2 \pi N^{1/\gamma}} e^{\log({b/\gamma})/\gamma} \gg \frac{g}{2 \pi N^{1/\gamma}},
   \ee
   where $b = O(1)$.  A similar result for the pairing scale has been obtained in Refs. ~\cite{max_last,max_private}
    using RG procedure.
Note in passing that the divergence of $T_c$ at $\gamma \to 0$ is the consequence of the fact that in this limit the effective interaction $\chi (\Omega_m) =(g/|\Omega_m|)^\gamma$ tends to a constant, while there is no upper cutoff in the theory. If we add a cutoff,
 we indeed obtain that
 $T_c$ saturates.

In  Fig. \ref{fig_2}  we plot  $T_c$ at $N=1$ obtained
   numerically from the Eliashberg equation (\ref{1}).
   We see that at  $\gamma >1$, $T_c$ rapidly approaches   $g/2\pi$ -- the result which we obtained analytically from the two lowest
     Matsubara frequencies.
   We emphasize that at both small and large $\gamma$ the
   fermionic self-energy is irrelevant for $T_c$.  At $\gamma \sim 1$, it does affect the value of $T_c$, but is not crucial in the sense that a comparable $T_c$ is obtained without including the self-energy.

  {\it \bf Conclusion.}~~~ In this paper we computed superconducting $T_c$ for a set of quantum-critical
  models with Eliashberg-type effective dynamical interaction between low-energy fermions.  We found that superconductivity always develops above
 a quantum-critical point, no matter what is the interplay between the pairing interaction and fermionic incoherence at $T=0$.
We argued that the proper calculation of $T_c$ should be  done directly  at a finite temperature, and
 $T_c$ is non-zero due to the fact that
 at a finite $T$  the self-energy vanishes at the two lowest fermionic Matsubara frequencies $\omega_m = \pm \pi T$.
   This implies that  fermionic incoherence at a QCP is not an obstacle for superconductivity.
     We caution, however, that this is true for the Eliashberg  $T_c$, which does not  include
      fluctuations of the pairing gap.
      The analysis of the gap fluctuations requires separate consideration.

 \acknowledgements
  We thank J. Carbotte,  R. Combescot, S. Hartnoll, G. Lonzarich, M. Metlitski, S. Kachru, S. Sachdev, G. Torroba, A-M Tremblay, H. Wang, and especially S. Raghu for useful discussions.   This work was supported in part by the NSF DMR-1523036 (AC),  the Gordon and Betty Moore Foundation's EPiQS Initiative through Grant No.\ GBMF4305 at the University of Illinois (YW), and by the David and Lucille Packard Foundation (EY).
Ar. A and A.C. are thankful to the Aspen Center for Physics where part of this work has been done. B.A., E.Y, and A.C. acknowledge partial support from KITP at UCSB. The research at KITP was supported in part by the National Science Foundation under Grant No. PHY11-25915.

\renewcommand{\theequation}{S\arabic{equation}}
\renewcommand{\thefigure}{S\arabic{figure}}
\renewcommand{\bibnumfmt}[1]{[S#1]}
\renewcommand{\citenumfont}[1]{S#1}

\onecolumngrid

\newpage
\centerline{\large{\bf Supplemental Material}}

\section{Cancellation of the singular self-action $m=m'$ term in Eq.\ (1)}

In the right hand side of Eq.\ (1) from the main text we have excluded the $m=m'$ terms for both the self-energy and the pairing vertex.
 We argued that this procedure is legitimate because such terms come from thermal fluctuations, which act  as non-magnetic impurity.
  In an $s$-wave superconductor,  the $m=m'$  contributions to the self energy and the pairing vertex  cancel out exactly, by Anderson's theorem.
    Here we show that singular contributions with $m=m'$ cancel out even if the pairing channel is different from an $s$-wave.

 We first emphasize that the $m =m'$ contribution has to be analyzed (and eliminated) within the original theory with $N=1$
 before  extending it to large $N$.  Otherwise, there will be unphysical divergencies. The elimination of the $m=m'$ term before large $N$ extension is legitimate
   because the goal of taking $N$ to be large is to enhance self-energy from quantum fluctuations [the one which leads to $\Sigma (\omega_m) \propto \omega^\alpha_m$ with $\alpha <1$]. The $m=m'$ term is not relevant at $T=0$, and its singular contribution comes from thermal fluctuations.  The extension to large $N$, which we used, is  tailored to analyze quantum, but not thermal fluctuations.

We
 next  remind the reader how the cancellation occurs for $s$-wave pairing.  For this, we consider the coupling
 via Einstein phonons
 (the case $\gamma =2$ in our notations).  The Eliashberg equations  are given by
\begin{align}
\Sigma( \omega_m)=& \pi T\sum_{m'} \chi(\omega_m-\omega_m')\sgn (\omega_m'),\nonumber\\
\Phi(\omega_m)=& \pi T\sum_{m'} \frac{\chi(\omega_m-\omega_m')}{|\omega_m'+\Sigma(\omega_m')|}\Phi(\omega_m'),
\label{s_eliash2}
\end{align}
where
\be
\chi(\omega_m-\omega_m')=\frac{g^2}{|\omega_m-\omega_m'|^2+\omega_E^2},
\ee
 is the propagator of an Einstein phonon.
In the limit  $\omega_E\to 0$, the $m=m'$ terms in both equations diverge. We keep $\omega_E$ finite at intermediate steps and set it to zero only at the end of calculations.

To see the cancellation of $m=m'$ terms, we introduce the gap function
\be
\Delta(\omega_m)\equiv\frac{\Phi(\omega_m)}{1+\Sigma(\omega_m)/\omega_m},
\ee
and re-express the Eliashberg equation for $\Phi (\omega_m)$ in \eqref{s_eliash2} as
\bea
\Delta(\omega_m)&=&T\sum_{m'}\chi(\omega_m-\omega_m')\[\frac{\Delta(\omega_m')}{\omega_m'}-\frac{\Delta(\omega_m)}{\omega_m}\] \nonumber\\
&& T\sum_{m' \neq m}\chi(\omega_m-\omega_m')\[\frac{\Delta(\omega_m')}{\omega_m'}-\frac{\Delta(\omega_m)}{\omega_m}\] \nonumber \\
&& + T \chi(0) \left[\frac{\Delta(\omega_m')}{\omega_m'}-\frac{\Delta(\omega_m)}{\omega_m}\right]_{m=m'}
\eea
 We see that the term with $m=m'$ vanishes, as long as $\omega_E$ is non-zero. Eliminating this term and re-introducing
\be
{\tilde \Sigma} ( \omega_m)=\pi T\sum_{m' \neq m } \chi(\omega_m-\omega_m')\sgn (\omega_m')
\ee
and
\be
{\tilde \Phi} (\omega_m) = \Delta(\omega_m)\left(1+\frac{{\tilde \Sigma}(\omega_m)}{\omega_m}\right),
\ee
we obtain the same set of Eliashberg equations as Eq. (\ref{s_eliash2}), but with $m' \neq m$ in the sum over Matsubara frequencies.
 Taking now the $\omega_E\to 0$ limit, we obtain Eq.\ (1) in the main text  for $\gamma=2$. [In the main text we  reverted to the un-tilded notation,  $\tilde\Sigma (\omega_m)\to\Sigma (\omega_m)$ and
 $\tilde\Phi (\omega_m)\to\Phi (\omega_m)$.  These variables, however, should not be confused with $\Sigma (\omega_m)$ and $\Phi(\omega_m)$ in Eq.~(\ref{s_eliash2}).

We now show that the  cancellation of the singular terms holds even when the pairing is not an $s$-wave.
   For this, we  move one step back and consider the pairing mediated by a critical collective mode with a momentum-dependent propagator $\chi(k_\|-k_\|',\omega_m-\omega_m')$
 between fermions at the Fermi surface ($|\bf k| = |{\bf k}'|=k_F$).

  For definiteness, we assume that the susceptibility is peaked at zero transferred momentum and set
\be
\chi({\bf k}-{\bf k'},\omega_m-\omega_m') \propto \frac {1}{({\bf k-k'})^2+ |\omega_m-\omega_m'|^{2\gamma}+\xi^{-2}}
\ee
At the critical point, $\xi^{-1} =0$, and 1D integration over ${k-k'}$ yields $\int dx \chi(x,\Omega_m) \propto 1/|\Omega_m|^\gamma$.  Other forms of $\chi$, which yield
the same  frequency dependence of the ``local'' susceptibility can also be used -- the end result of the analysis of $m =m'$ term will be the same.

Integrating in the fermionic propagators over the momenta transverse to the Fermi surface, we obtain the set of Eliashberg equations in the form
\begin{align}
\Sigma(k_\|, \omega_m)=&T\sum_{m'}\int \frac{dk'_\|}{4\pi v_F} \chi(k_\|-k_\|',\omega_m-\omega_m')\sgn (\omega_m'),\nonumber\\
\Phi(k_\|,\omega_m)=&T\sum_{m'} \int \frac{dk'_\|}{4\pi v_F}\frac{\chi(k_\|-k_\|',\omega_m-\omega_m')}{|\omega_m'+\Sigma(k_\|',\omega_m')|}\Phi(k_\|',\omega_m'),
\label{s_eliash}
\end{align}
where both ${\bf k}$ and ${\bf k}'$ are on the Fermi surface.
In both equations in \eqref{s_eliash}, the integrals over $k_\|'$ in the term with $m=m'$ are singular at the quantum critical point $\xi^{-1}=0$.
To regularize the divergence, we again keep $\xi$ finite at intermediate steps and set it to infinity only at the end of calculations.

\begin{figure}
\includegraphics[width=\columnwidth]{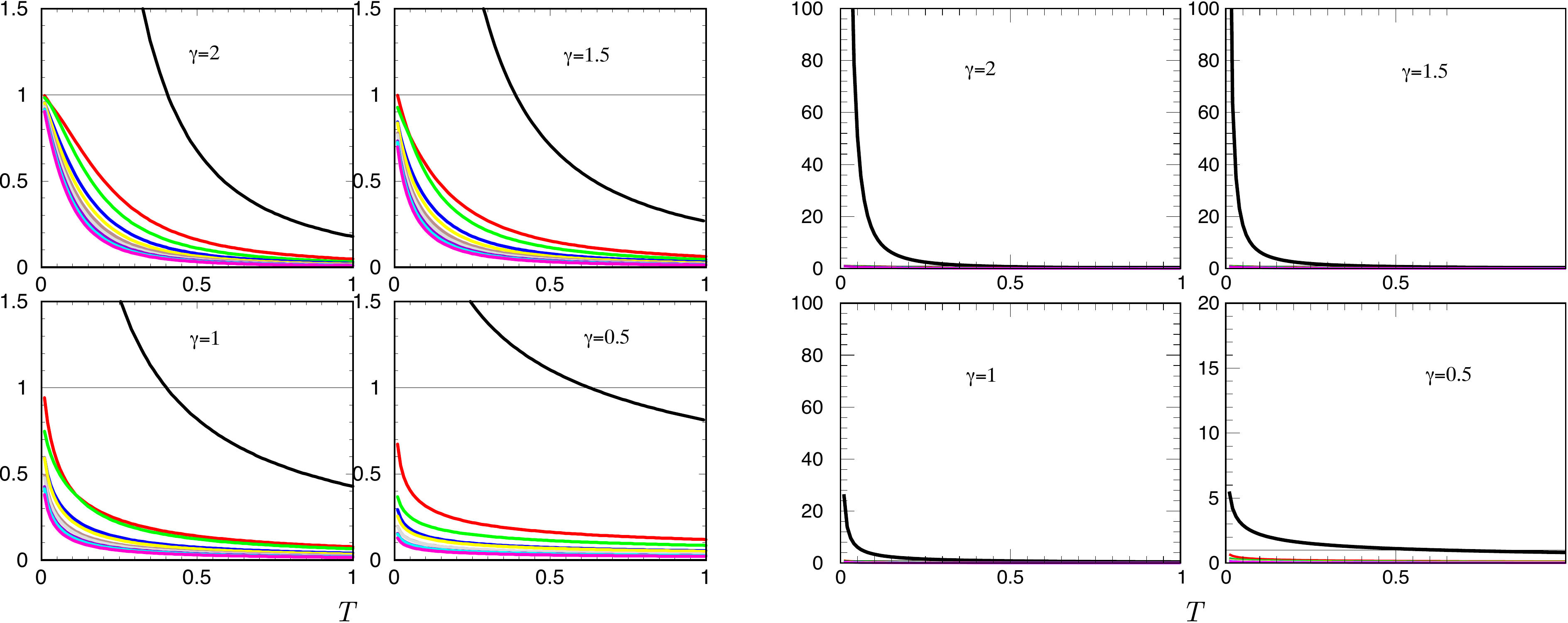}
\caption{ The first 10 eigenvalues of the equation for $\Phi (\omega_m)$ for different $\gamma$.
 The existence of the threshold in $N$ would imply that the largest eigenvalue saturates at a finite value at $T \to 0$. We see instead that it keeps increasing with decreasing $T$. We verified that the largest eigenvalue $\lambda$  follows our result: $\lambda \propto 1/T^\gamma$. Upper and lower panels show the same behavior in different vertical ranges.}
\label{s_evs_vs_temp}
\end{figure}

Like before, we introduce the superconducting gap function as
\be
\Delta(k_\|,\omega_m)\equiv \frac{\Phi(k_\|,\omega_m)}{1+\Sigma(k_\|,\omega_m)/\omega_m},
\ee
and reexpress the equation for $\Phi$ in \eqref{s_eliash} as
\bea
\Delta(k_\|,\omega_m)=&& T\sum_{m'} \int \frac{dk_{\|}'}{4\pi} \chi(k_\|-k_\|',\omega_m-\omega_m')\[\frac{\Delta(k_\|',\omega_m')}{|\omega_m'|}-\frac{\Delta(k_\|,\omega_m)}{|\omega_m|}\] \nonumber \\
 = &&T\sum_{m' \neq m} \int \frac{dk_{\|}'}{4\pi} \chi(k_\|-k_\|',\omega_m-\omega_m')\[\frac{\Delta(k_\|',\omega_m')}{|\omega_m'|}-\frac{\Delta(k_\|,\omega_m)}{|\omega_m|}\] \nonumber \\
&& + T \int \frac{dk_{\|}'}{4\pi} \chi(k_\|-k_\|') \frac{\Delta(k_\|',\omega_m)- \Delta(k_\|,\omega_m)}{|\omega_m|}
\label{s_gap}
\eea

    The terms with $\Delta(k_\|',\omega_m)$ and $\Delta(k_\|,\omega_m)$ in the last line in Eq. (\ref{s_gap}), when taken separately,  diverge at $\xi \to \infty$.
    However, the singular terms obviously cancel out in the full expression.  This has been noted in Ref.\ \onlinecite{s_msv}.
     The remainder is non-singular, but is also non-zero for a $k$-dependent gap function, and we label it as
    \be
    \Delta(k_\|,\omega_m) \beta(k_\|,\omega_m)\equiv T \int \frac{dk_{\|}'}{4\pi} \chi(k_\|-k_\|') \frac{\Delta(k_\|',\omega_m)- \Delta(k_\|,\omega_m)}{|\omega_m|}.
    \ee
      Introducing then
    \be
 \tilde \Sigma(k_\|, \omega_m)=T\sum_{m' \neq m}\int \frac{dk'_\|}{4\pi v_F} \chi(k_\|-k_\|',\omega_m-\omega_m')\sgn (\omega_m'),
    \ee
     and
 \be
{\tilde \Phi} (k_\|,\omega_m) =  \Delta(k_\|,\omega_m) \left(1+\frac{{\tilde \Sigma}(k_\|,\omega_m)}{\omega_m}\right),
\ee
we re-express Eliashberg equations as
\begin{align}
{\tilde \Sigma} (k_\|, \omega_m)=&{\tilde \Sigma} (\omega_m) = T\sum_{m' \neq m}\int \frac{dk'_\|}{4\pi v_F} \chi(k_\|-k_\|',\omega_m-\omega_m')\sgn (\omega_m'),\nonumber\\
{\tilde \Phi(k_\|,\omega_m)} \left(1 - \frac{\beta(k_\|,\omega_m)}{1 +{{\tilde \Sigma(k_\|,\omega_m)}}/{\omega_m}}\right) =&T\sum_{m' \neq m} \int \frac{dk'_\|}{4\pi v_F}\frac{\chi(k_\|-k_\|',\omega_m-\omega_m')}{|\omega_m'+\Sigma(k_\|',\omega_m')|}\tilde\Phi(k_\|',\omega_m'),
\label{s_eliash_a}
\end{align}
In a general case, this expression cannot be reduced to integral equation in frequency only, because $\beta$ term cannot be absorbed into the coupling constant, and also because
 the integration over $k'$ along the Fermi surface cannot be carried out explicitly if $\Delta$ is momentum-dependent.  This last complication is often by-passed by
  approximating the momentum dependence of $\Delta$ by one of lattice harmonics. Then momentum integral can  be done explicitly.

  An additional complication arises when $\chi (k, \omega_m)$ is peaked at finite ${\bf k}$ [${\bf k} = (\pi,\pi)$ for antiferromagnetic QCP].
      Then integration over  the component of ${\bf k}'$ transverse to the Fermi surface necessarily  introduces an additional term into the bosonic propagator for ${\bf k} = {\bf k}' = k_F$, when ${\bf k}_F$ is away from a hot spot.  As a result,  the momentum integration in  Eq. (\ref{s_eliash_a}) cannot be carried out explicitly
       even if one approximates the gap functions in hot regions as constants and neglect the dependence of the self-energy  ${\tilde \Sigma} (k_\|', \omega_m)$
         on $k_\|'$.  For example, if we take antiferromagnetic spin-fermion model with
  \be
\chi(k,\Omega_m)=\frac {{\bar g}}{k^2+ \gamma |\Omega_m| +\xi^{-2}},
\ee
 expand the dispersion near the hot spot ${\bf k}_h$ as $\epsilon_k = v_x {\delta k}_x + v_y {\delta k}_y$ and the one near the hot spot at ${\bf k}_h + (\pi,\pi)$ as
  $\epsilon_{k + (\pi,\pi)} = v_x {\delta k}_x - v_y {\delta k}_y$,  integrate over momenta $\bf k'$ in the gap equation transverse to the Fermi surface and approximate the bosonic propagator by its value for particles at the Fermi surface, we obtain effective
  \be
  \chi (k_\parallel, k_\parallel', \omega_m) = \frac{\bar g}{(k'_{\parallel}-k_\parallel)^2 + [4v^2_y/(v^2_x + v^2_y)] k^2_\parallel + \gamma |\Omega|_m + \xi^{-2}}.
  \label{s_s16}
  \ee
    Without the additional $k^2_\parallel$ term one could explicitly integrate over $k_\parallel - k'_\parallel$ and obtain Eq.\ (1) from the main text with $g = {\bar g}^2/16 \gamma$.  In the presence of the  $k_\|^2$ term, the kernel is no longer a function of $k'_\parallel - k_\parallel$, and one has to solve integral equation in both momentum and frequency.

  This additional complication, however, does not introduce qualitative changes as additional terms do not change the scaling form of the gap equation at relevant frequencies
  (i.e., typical $k'_\parallel \sim k_\parallel \sim \gamma \Omega_m$).
 Furthermore, the strength of the additional term in Eq.\ \eqref{s_s16} for the antiferromagnetic case and the strength of $\beta$ term,  which is the ``leftover" from $m = m'$ term in the gap equation for non-$s$-wave gap, depend  on the ratio of $v_y/v_x$ (Ref.\ \onlinecite{s_efetov}) and  can be made small by varying the geometry of
 the Fermi surface.
    For this reason, we neglected the complications due to the ``leftover" term $m=m'$ and due to the presence of additional $k^2_\parallel$ in $\chi (k, k', \Omega_m)$  in the hot spot case (and to momentum dependence of the self-energy along the Fermi surface), integrated explicitly over $k'_\parallel - k_\parallel$, and analyzed the integral equation for $\Phi (\omega_m)$ in the frequency domain.

\section{The failure of the argument for the threshold}

We present numerical and analytical arguments for the failure of the argument for the existence of the threshold in $N$ for superconductivity at the quantum-critical point.

 The numerical argument is the following.  If the threshold exists, then at the threshold value $N = N_{cr}$, the linearized gap equation has a solution at $T=0$. One can verify this by solving the gap equation as an eigenvalue/eigenfunction equation, with $N$ playing the role of the eigenvalue. The existence of $N = N_{cr}$ would imply that the largest eigenvalue of the gap equation tends to a finite value $N_{cr}$ at $T \to 0$.

 In Fig. \ref{s_evs_vs_temp} we present the solution of the gap equation for different $\gamma$ on a mesh of $2000 \times 2000$ Matsubara frequencies.
  We clearly see that for all $\gamma$, one eigenvalue gradually increases as $T$ decreases and shows  no tendency of saturation. We have explicitly verified that at small $T$, this eigenvalue grows as $1/T^\gamma$, precisely as we found in the main text.

 There are several analytical arguments as well.  They all are valid for $\gamma <1$, when there is a finite frequency $\omega_0$, above which
  the self-energy $\Sigma (\omega_m) = (\omega_m)^{1-\gamma} \omega^\gamma_0$ becomes smaller than the bare frequency.  The analysis of the gap equation at $T=0$ for $\gamma >1$ is more involved as it requires a proper regularization of divergencies.  There is no such problem in finite $T$ analysis because the self-energy is finite at $T >0$ for all $\gamma$.

   The goal of analytical reasoning is to  show that the oscillatory solution with $\chi_{pp} (\omega_m) \propto (\omega_0/|\omega_m|)^{\gamma/2 \pm i\beta}$, which, we remind,  is valid for $\omega_m \ll \omega_0$, cannot be matched with the solution at $\omega_m \gg \omega_0$.  One argument, which requires some lengthy calculations and will be presented in a separate publication, is based on the fact that, by power counting, the corrections to the oscillatory solutions  should hold in powers of $(\omega_m/\omega_0)^\gamma$. These corrections come from keeping the bare $\omega$ term in the fermionic propagator along with the fermionic self-energy $\Sigma (\omega_m)$.  If the prefactors for the series were finite, the corrections would be determined by internal $\omega^{'}_m$ of order $\omega_m$, i.e. the theory would remain local. The locality is essential as it is assumed when the pairing problem is reformulated in terms of a differential RG equation rather than  the integral equation, as in Refs.\ \onlinecite{s_11,s_14,s_16}.
 However, the calculation shows that the prefactors for these corrections are ultra-violent divergent within the perturbation theory, which implies that they are determined by internal frequencies $\omega^{'}_m$ of order $\omega_0$ rather than $\omega_m$.  This implies that beyond leading order in $\omega$, the theory becomes non-local,
   and the exact behavior even at the smallest frequencies is determined by frequencies of order $\omega_0$, where the bare $\omega$ term in the fermionic propagator is no longer small compared to $\Sigma (\omega_m)$.

   Another analytical argument against the existence of the threshold is the following.
   Consider again the  equation for $\Phi (\omega)$ at $T=0$ (Eq. (1) from the main text), and rescale frequencies by $g$.  We then have
\begin{equation}\label{s_eq:eq}
\Phi (\Omega )=\frac{1-\gamma}{2N}\int_{-\infty }^{\infty } \frac{\Phi (\omega )d\omega }{|\omega -\Omega |^{\gamma }|\omega|^{1-\gamma }}\frac{1}{1+(1-\gamma )|\omega|^{\gamma }}
\end{equation}
Let's define the oscillatory solutions $\Phi (\omega_m) \propto |\omega_m|^{-\gamma/2 \pm  i {\bar \beta}}$ as  $\Phi_{\pm }(\Omega )$. These $\Phi_{\pm }(\Omega )$ satisfy
\begin{equation}\label{s_eq:eq_1}
\Phi_{\pm} (\Omega )=\frac{1-\gamma}{2N}\int_{-\infty }^{\infty } \frac{\Phi (\omega )d\omega }{|\omega -\Omega |^{\gamma }|\omega|^{1-\gamma }}
\end{equation}
Let's multiply  Eq.~\eqref{s_eq:eq}  by $\frac{\Phi_{\pm }(\Omega )}{|\Omega |^{1-\gamma }}$ and integrate it over $\Omega $. We obtain
\be
\int_{-\infty}^{\infty }\frac{\Phi (\Omega )\Phi_{\pm }(\Omega )d\Omega} {|\Omega |^{1-\gamma }}=\frac{1-\gamma }{2N}\int_{-\infty}^{\infty }\frac{\Phi (\omega )d\omega }{|\omega |^{1-\gamma }\left(1+(1- \gamma )|\omega |^{\gamma } \right)}\int_{-\infty}^{\infty } \frac{\Phi (\Omega )d\Omega }{|\omega -\Omega |^{\gamma }|\Omega|^{1-\gamma }} =\int_{-\infty}^{\infty }\frac{\Phi (\omega )\Phi_{\pm}(\omega )d\omega }{|\omega |^{1-\gamma }}\frac{1}{1+(1- \gamma )|\omega |^{\gamma }}
\ee
We see that if  Eq.~\eqref{s_eq:eq} has a solution $\Phi (\omega )$, then it must satisfy
\be
\label{s_eq:G1}
 \int_{-\infty}^{\infty }\frac{\Phi (\omega )G_{1}(\omega )d\omega}{|\omega |^{1-\gamma }\left(1+(1- \gamma )|\omega |^{\gamma } \right)} =0,
 \ee
 where
 \be
 \qquad G_{1}(\omega )=|\omega |^{\gamma }\Phi_{\pm}(\omega ) = |\omega |^{\gamma /2\pm i{\bar \beta} }
\ee
Now  multiply Eq.~\eqref{s_eq:eq}  by $\frac{G_{1}(\Omega )}{|\Omega |^{1-\gamma }\left(1+(1- \gamma )|\Omega |^{\gamma } \right)} $ and integrate over $\Omega $ to obtain
\be
0=\int_{-\infty}^{\infty }\frac{\Phi (\omega )d\omega }{|\omega |^{1-\gamma }}\frac{1}{1+(1-\gamma )|\omega|^{\gamma }}
\int_{-\infty}^{\infty } \frac{d\Omega }{|\Omega -\omega |^{\gamma }}\frac{G_{1}(\Omega )}{|\Omega |^{1-\gamma }\left(1+(1- \gamma )|\Omega |^{\gamma } \right)}  =\int_{-\infty}^{\infty }\frac{\Phi (\omega )G_{2}(\omega )d\omega}{|\omega |^{1-\gamma }\left(1+(1- \gamma )|\omega |^{\gamma } \right)},
\ee
where
\begin{equation}\label{s_eq:G2}
G_{2}(\omega )=\int_{-\infty}^{\infty } \frac{d\Omega }{|\Omega -\omega |^{\gamma }}\frac{G_{1}(\Omega )}{|\Omega |^{1-\gamma }\left(1+(1- \gamma )|\Omega |^{\gamma } \right)},
\end{equation}
and so on. We will get then the series of functions
\begin{equation}\label{s_eq:Gn}
G_{n}(\omega )=\int_{-\infty}^{\infty } \frac{d\Omega }{|\Omega -\omega |^{\gamma }}\frac{G_{n-1}(\Omega )}{|\Omega |^{1-\gamma }\left(1+(1- \gamma )|\Omega |^{\gamma } \right)},
\end{equation}
such that for any $n$,
\be
\int_{-\infty}^{\infty }\frac{\Phi (\omega )G_{n}(\omega )d\omega}{|\omega |^{1-\gamma }\left(1+(1- \gamma )|\omega |^{\gamma } \right)}
=0.
\ee
At small frequencies, the oscillatory solution is $\Phi (\omega_m) = |A| \left( e^{i\phi} \Phi_{+} (\omega_m) + e^{-i\phi} \Phi_{-} (\omega_m)\right)$
 This solution has a single free parameter -- a phase $\phi$. One parameter cannot satisfy infinite set of equation, unless all $G_n$ are multiples of $G_1$. One can easily check explicitly that they are not.   This implies that the oscillatory solution is actually not the solution of the actual equation for $\Phi (\omega_m)$, even at the smallest frequencies.

\section{$T_c$ at large $N$}
We begin with Eq.\ (2) of the main text
\begin{align}
{\bar \Phi}_0 =&\frac{K_T}{N} {\bar \Phi}_{-1} +\frac{1}{N} \sum_{m>0} \frac{{\bar \Phi}_m}{{\bar \Sigma}_m} \left[\frac{1}{m^\gamma} + \frac{1}{(m+1)^\gamma}\right],   \nonumber \\
{\bar \Phi}_{m>0} =&  \frac{K_T}{N}  \left[\frac{{\bar \Phi}_0}{m^\gamma} + \frac{{\bar \Phi}_{-1}}{(m+1)^\gamma}\right]
+\frac{1}{N} \sum_{m'>0, m' \neq m} \frac{{\bar \Phi}_m}{{\bar \Sigma}_m} \left[\frac{1}{|m-m'|^\gamma} + \frac{1}{(m+m'+1)^\gamma}\right] \label{s_4},
\end{align}
where we remind $\bar\Phi_m=\Phi_{m}/(\pi T K_T)$, $K_T=[g/(2\pi T)]^\gamma$, and
\be
\bar\Sigma_m=\sum_{m'\neq m}\frac{\sgn(2m'+1)}{|m-m'|^\gamma}.
\label{s_eq7}
\ee
At the leading order in $1/N$, only the first term in the r.h.s. of the equation for  ${\bar \Phi}_0$ matters, and using
 $\bar\Phi_0\equiv \bar\Phi_{-1}$, we immediately obtain that $T_c$ is the solution of  $K_T/N=1$, i.e.,
  $T_c\approx {g}/(2\pi N^{1/\gamma})$.

  To obtain the next order correction to $T_c$, we rewrite the first equation of (\ref{s_4}) as
\be
\frac{K_T}{N}=1-\frac{1}{N}\sum_{m>0}\frac{\tilde \Phi_m}{\bar\Sigma_m}\[\frac{1}{m^\gamma}+\frac{1}{(m+1)^\gamma}\],
\label{s_5}
\ee
where $\tilde\Phi_m\equiv \bar\Phi_m/\bar\Phi_0$, and its leading order value is, from second equation of (\ref{s_4}),
\be
\tilde\Phi_m=\[\frac{1}{m^\gamma}+\frac{1}{(m+1)^\gamma}\].
\label{s_6}
\ee
Plugging (\ref{s_6}) into (\ref{s_5}) we obtain
\be
\frac{K_T}{N}=1-\frac{1}{N}\sum_{m>0}\frac{1}{\bar\Sigma_m}\[\frac{1}{m^\gamma}+\frac{1}{(m+1)^\gamma}\]^2
\equiv 1-\frac{\delta_\gamma}{N}.
\ee
Solving this equation for $T_c$, we obtain at a large $N$
\be
T_c\approx \frac{g}{2\pi}\frac{1}{N^{1/\gamma}}\(1+\frac{\delta_\gamma}{N\gamma}\),
\ee
 where $\delta_\gamma=\sum_{m>0}{[1/m^\gamma+1/(m+1)^\gamma]^2}/\bar\Sigma_m$.
 This is Eq.\ (3) from the main text.  At $\gamma\to \infty$, only the first term in the sum contributes to $\delta_\gamma$. Using the fact that $\bar\Sigma_1=2$ [see Eq.\ (\ref{s_eq7})], we obtain $\lim_{\gamma\to\infty} \delta_\gamma=1/2$.

\section{$T_{c}$ at large $\gamma $}\label{s_sec:large}
We again depart from Eq. (1) in the main text
\begin{eqnarray}
     &&\Phi (\omega_m) = \frac{1}{N} g^\gamma \pi T \sum_{m' \neq m} \frac{\Phi (\omega_{m'})}{|\omega_{m'} + \Sigma (\omega_{m'})|} ~\frac{1}{|\omega_m - \omega_{m'}|^\gamma} \nonumber \\
    && \Sigma (\omega_m) = g^\gamma \pi T \sum_{m' \neq m}  \frac{{\text{sign}}(\omega_{m'})}{|\omega_m - \omega_{m'}|^\gamma}
\label{s_1}
\end{eqnarray}

In the limit $\gamma \rightarrow \infty $ we neglect all terms in the sums except the ones with $m'=m\pm 1 $. Then after the change  of notations
\be
\phi_{m}=  \frac{\Phi (\omega_{m'})}{|\omega_{m'} + \Sigma (\omega_{m'})|},
\ee
we get
\begin{eqnarray}
&&\phi_{m} |\omega_{m} + \Sigma_{m}|=\frac{1}{N} \pi K_{T} T(\phi_{m-1}+\phi_{m+1}),\nonumber\\
&&\Sigma_{m}=\left\{ \begin{array}{ll}
0,&m=0,-1;\\
2\pi K_{T}T,&m>0;
\end{array}\right.
\end{eqnarray}
where we used the notation introduced in the paper $K_{T}=\left(\tfrac{\bar{g}}{2\pi T} \right)^{\gamma }$.

Treating $m=0$ and $m>0$ separately we find
\begin{equation} \label{s_eq:recurrence}
\phi_{1}=\phi_{0}\left(s -1\right),\qquad \phi_{m+1}= \phi_{m} \left( s (2m+1) + 2N\right)-\phi_{m-1},
\end{equation}
where $s=N/K_{T}$.
We see, that for a  given $\phi_{0}$,
  this equation yields all $\phi_{m}$ for $m>1$.  In addition, we need to satisfy the first relation in \eqref{s_eq:recurrence} and the condition $\phi_{m\rightarrow \infty }\rightarrow 0$.

The recurrence relation \eqref{s_eq:recurrence} can be solved by the method of generating functions.  We also   recognize in \eqref{s_eq:recurrence} the recurrence relation for the Bessel functions
$zZ_{\nu -1}(z)+zZ_{\nu +1}=2\nu Z_{\nu }(z)$, where $Z_{\nu }(z)$ is any of Bessel functions.

The Bessel function, which tends to $0$ as $m\rightarrow \infty $, is $J_{\nu }(z)$, so
\be
 \phi_{m}= aJ_{m+1/2+N/s}(1/s),
\ee
where $a$ is an arbitrary constant.  We now need to satisfy  the first relation of \eqref{s_eq:recurrence}, i.e.
\be
  aJ_{3/2+N/s}(1/s)=(s-1)aJ_{1/2+N/s}(1/s).
\ee
Thus
\begin{equation}\label{s_eq:gcVsN}
\frac{J_{3/2+N/s}(1/s)}{J_{1/2+N/s}(1/s)}=s-1.
\end{equation}
This equation defines $s$ (or $K_{T}$) as a function of $N$.
\begin{figure}
\includegraphics[width=0.5\columnwidth]{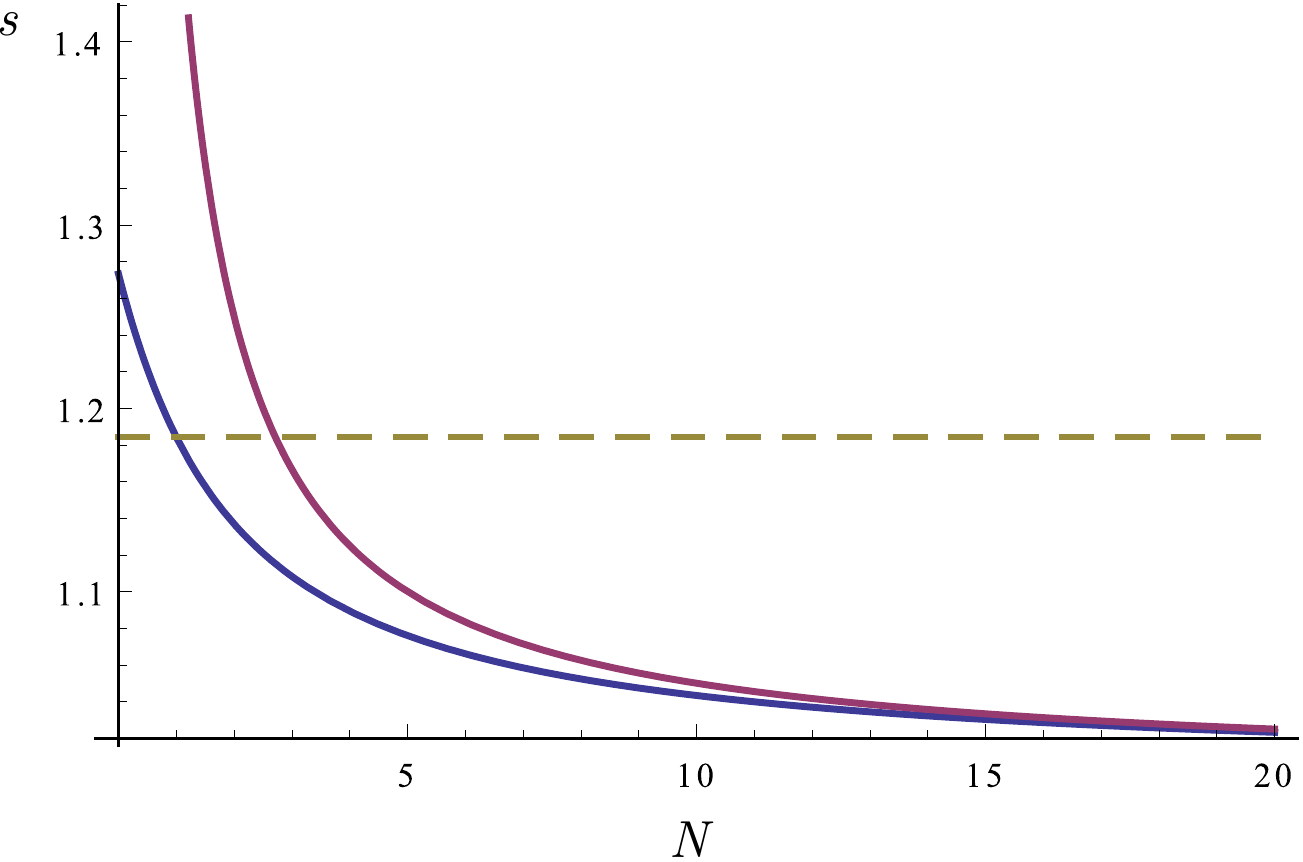}
\caption{\label{s_fig:TcVsN} $s$ vs. $N$ given by Eq.~(\ref{s_fig:TcVsN}) -- blue line, $1+\frac{1}{2N}$ -- red line, $s(N=1)$ -- yellow dashed line.
}
\end{figure}

Solving Eq.~\eqref{s_eq:gcVsN}  numerically    for $N=1$  we obtain
\be
s(N=1)  = 1.1843,
\ee
as we stated in the main text.
At large $N$, using the asymptotic expansion of the Bessel function for large order at fixed argument,
$$
J_\nu(z)=\frac{1}{\sqrt{2\pi\nu}}\left( \frac{e z}{2\nu}\right)^\nu\!\!\!\!,
$$
we find
\be
s\approx 1+\frac{1}{2N}.
\ee
The full function  $s(N)$ is shown in Fig. \ref{s_fig:TcVsN}.

\section{$T_{c}$ at small  $\gamma $}\label{s_sec:small}
 
 Here we show the details of our computation of  $T_c$ at small $\gamma$ and $N = O(1)$.
 Fig. 3 from the main text shows that in this regime the actual $T_c$ is much larger than the contribution of only the lowest Matsubara frequencies, i.e. it is determined by
  $\omega_m$ with $m >0$.
   This figure also
    shows that
  $T_c \gg \omega_0$, where  $\omega_0= g [2/(1-\gamma)]^{1/\gamma}$  is the scale below which the self-energy $\Sigma (\omega_m) = (\omega_m)^{1-\gamma} \omega^\gamma_0$ is larger than $\omega_m$.
  Then, the self-energy is irrelevant  simply because at $T= T_c$, $\omega_m \gg \Sigma (\omega_m)$  for all $m$.
 Neglecting $\Sigma (\omega_{m'})$ in
  Eq. (\ref{s_1}),
    replacing the summation over $m$ by the integration over the frequency (which is justified because typical $m' \gg 1$),
    introducing new variable $x = (\omega_0/\omega)^\gamma$ and assuming that $\Phi (x)$ is a smooth function, we reduce integral equation (\ref{s_1}) to a differential equation
\be
   \Phi{''} (x)  + \frac{\Phi(x)}{x \gamma N} =0,
\ee  
  with the boundary condition $\Phi (x \ll 1) \propto x$   (i.e. $\Phi (\omega_m) \propto 1/|\omega_m|^\gamma$ at $\omega_m \gg \omega_0$).   The solution of this equation is  
  \be
  \Phi (x) \propto \sqrt{x} J_1 \left(2 \sqrt{x/\gamma N}\right),
  \ee
   where $J_1$ is a Bessel function of the first kind.
   At  small argument $J_1 (y) \propto y$, at large $y$   $J_1 (y)$  oscillates.   Because $x \sim \gamma N$ is the only scale in $\Phi (x)$, it is natural to identify the corresponding frequency $\omega_m = \omega_0/x^{1/\gamma}$ with $T_c$.  This yields
  \be
  T_c \sim \omega_0(\gamma N)^{-1/\gamma}  \sim  \frac{g}{2 \pi N^{1/\gamma}} e^{\log({b/\gamma})/\gamma} \gg \frac{g}{2 \pi N^{1/\gamma}},
  \ee
   where $b = O(1)$.

\end{document}